\documentclass[baaa]{baaa}

\usepackage[pdftex]{hyperref}
\usepackage{subfigure}
\usepackage{natbib}
\usepackage{helvet,soul}
\usepackage[font=small]{caption}

\contriblanguage{1}

\contribtype{1}

\thematicarea{6}

\title{
A model for molecular hydrogen-dependent star formation
 in simulations of galaxy evolution}

\titlerunning{A model for molecular hydrogen-dependent star formation}

\author{
E. Lozano\inst{1,2}, C. Scannapieco\inst{1,2} \& S.E. Nuza\inst{2,3}
}

\contact{elozano@df.uba.ar}

\institute{
Departamento de Física, Facultad de Ciencias Exactas y Naturales, UBA, Argentina
\and
Consejo Nacional de Investigaciones Científicas y Técnicas, Argentina
\and
Instituto de Astronomía y Física del Espacio, CONICET--UBA, Argentina
}

\resumen{
La formación estelar, junto con el {\it feedback} químico y energético asociado, es uno de los procesos más importantes en la evolución de las galaxias. La actividad de formación de estrellas en galaxias define y afecta muchas de sus propiedades fundamentales, como la masa estelar, la morfología y el nivel de enriquecimiento químico. Los modelos simples para la formación de estrellas en simulaciones hidrodinámicas cosmológicas han demostrado ser exitosos al reproducir los niveles de tasa de formación estelar (SFR) así como las formas típicas para distintos tipos de galaxias. Sin embargo, con el advenimiento de simulaciones de alta resolución y observaciones más detalladas, es necesario contar con modelos de formación estelar más sofisticados; en particular, para comprender mejor la relación entre la SFR  y la cantidad de gas en las fases atómica y molecular. En este trabajo, aplicamos un nuevo modelo de formación estelar, recientemente desarrollado para trabajar en el contexto de simulaciones hidrodinámicas, al estudio de la SFR en galaxias de masa similar a la Vía Láctea. El nuevo modelo describe la formación de hidrógeno molecular a partir de material atómico, considerando también posibles dependencias con la abundancia química del gas. Esto permite implementar varios modelos de formación estelar, donde la SFR de una nube de gas depende de las fracciones atómica y/o molecular, y comparar sus predicciones con resultados observacionales.
}

\abstract{
Star formation, together with the associated chemical and energy feedback, is one of the most important processes in galaxy evolution. The star formation activity in galaxies defines and affects many of their fundamental properties, such as stellar mass, morphology and chemical enrichment levels. Simple models for star formation in cosmological hydrodynamical simulations have shown to be successful in reproducing the star formation rate (SFR) levels and shapes of different types of galaxies. However, with the advent of high-resolution simulations and more detailed observations, more sophisticated star formation models are needed; in particular, to better understand the relation between star formation and the amount of gas in the atomic and molecular phases. In this work, we apply a novel star formation model, recently developed to work in the context of hydrodynamical simulations, to the study of the SFR in Milky Way-mass galaxies. The new implementation describes the formation of molecular hydrogen from atomic material, considering also possible dependencies with the chemical abundance of the gas. This allows to implement various star formation models, where the SFR of a gas cloud is determined by the atomic and/or molecular gas phases, and to compare their predictions to recent observational results.
}


\begin{document}

\maketitle

\section{Introduction}
\label{sec:intro}

The star formation rate (SFR) is a key characteristic of galaxies. In the context of the standard cosmological model, the SFR is determined by a combination of various processes that take place over the course of a galaxy's lifetime, such as gas cooling, star formation, chemical enrichment, and feedback from supernovae and galactic nuclei. These processes are influenced by factors like mergers, interactions, and mass accretion, which affect the amount and properties of the gas from which stars form. The density of a gas cloud is believed to be the most important factor in determining its star formation rate, although the details of this process are not yet fully understood.
Observationally, the total gas density is found to be correlated to the star formation rate \citep{Kennicutt1998}, and this correlation is even stronger when considering the molecular gas \citep{Wong2002,Bigiel2008}.

As the formation of dark matter halos and galaxies is highly non-linear, numerical simulations have become the preferred tool to investigate how galaxies form and evolve from early times up to the present.
This type of simulations naturally include mergers/interactions and continuous gas accretion, processes that may induce changes in the SFR. However, there are still significant uncertainties in the modelling of the evolution of the baryonic component, since the physical processes that affect baryons – such as star formation, feedback, and chemical enrichment – take place at scales that are too small to be resolved directly. As a result, these processes are  introduced using sub-grid physics, which involves a number of adjustable parameters that are not always independent of one another or constrained observationally. This can lead to inconsistencies in the predictions of different models \citep{Scannapieco2012}.

Because of its importance in galaxy formation, it is critical for simulations to accurately describe the star formation process at the scales that can be resolved, as well as the associated feedback effects. In this work, we present a new model of star formation that takes into account the chemical composition and the relative abundance of atomic and molecular gas phases. The model is grafted onto the cosmological, magnetohydrodynamical code {\sc Arepo} \citep{Springel2010}.

The remainder of this work is organized as follows: in Sec.~\ref{sec:model}, we describe our new model; in Sec.~\ref{sec:results}, we discuss the results for simulations of an isolated galaxy; and in Sec.~\ref{sec:conclusions}, we present our conclusions. 

\section{The star formation model}
\label{sec:model}

Our star formation model is designed to track the time evolution of the molecular and atomic  phases of hydrogen in a gas cloud \citep{Ascasibar2015,Molla2017,Millan-Irigoyen2020}, and be applied to numerical simulations of the formation of galaxies. By following the evolution of these two phases separately, we can construct different prescriptions for a star formation law linked to the amount of molecular and atomic gas in different proportions.
In this way, we improve the traditional star formation law where the star formation rate is only a function of the total gas density and does not depend explicitly on the molecular or atomic fractions \citep{Katz1992}.

The evolution of the neutral gas and stellar components in a gas cell is obtained by solving a system of coupled equations for the atomic ($a_f$), molecular ($m_f$), and (newly formed) star ($s_f$) fractions, namely:

\begin{align}
    & \dot{a}_f(t)= -a_f(t) \, \tau_C^{\,-1} + (\eta + R) \, \psi(t) \, , \\
    & \dot{m}_f(t)= a_f(t) \, \tau_C^{\,-1} - (\eta + 1) \, \psi(t) \, , \\
    & \dot{s}_f(t)= (1 - R) \, \psi(t) \, .
\end{align}

The exchange of mass between phases is driven by dissociation of molecular hydrogen -- with an efficiency per unit of star formation rate $\eta$ -- and condensation of atomic hydrogen -- catalyzed by dust grains \citep{Millan-Irigoyen2020} and regulated by the time parameter $\tau_C$. 
For the star formation rate we use the following parametrization:
\begin{equation}
    \psi(t) = \frac{m_w \, m_f(t) + a_w \, a_f(t)}{\tau_S} \, ,
\end{equation}
where $m_w$ and $a_w$ are the relative weights of the atomic and molecular fractions (which can vary for different models) and $\tau_S$ is a typical time-scale. 

\begin{figure}[!ht]
    \centering
    \includegraphics[width=0.8\columnwidth]{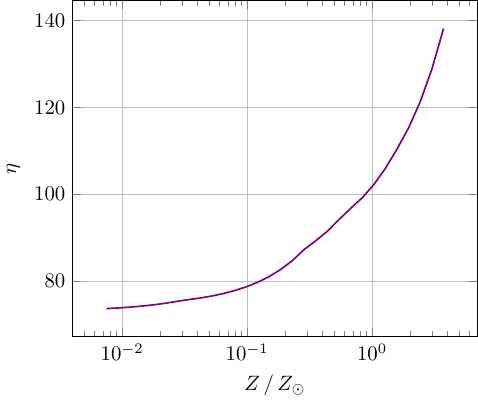}
    \caption{Dependence of the dissociation of molecular hydrogen parameter $\eta$ with metallicity for the IMF of \cite{Chabrier2003} and an integration time of $1 \, \mathrm{Myr}$.}
    \label{fig:eta_vs_z}
\end{figure}

The mass return from stars to interstellar gas due to supernovae is characterized by the parameter $R$, and assumed to be in the atomic phase. This assumption is based on the instantaneous recombination hypothesis, which states that the ionized hydrogen transforms into atomic hydrogen in a timescale much shorter than the integration time. This allows us to accurately model the mass return and its effects on the interstellar gas.

Most of the input parameters of the model are constants that are well constrained empirically or theoretically, that we take from the literature. In contrast, $\tau_C$ depends on the properties of the gas, particularly the gas metallicity ($\propto 1/Z$) and density ($\propto 1/\rho$), and these dependencies are considered in our model following \cite{Osterbrock2006} and \cite{Millan-Irigoyen2020}.
Furthermore, $\eta$ depends on the  metallicity of the newly formed stars -- assumed to be inherited from the gas --  and the integration time step; Fig.~\ref{fig:eta_vs_z} shows the relation between $\eta$ and $Z$ for a time-step of $1\,\mathrm{Myr}$.

\section{Results}
\label{sec:results}

We run several simulations using an idealized initial condition for a Milky Way-mass halo with a virial mass of $10^{12} \, \mathrm{M_\odot}$, producing a galaxy with spiral morphology. We assumed different values for the input parameters which link the star formation to the atomic/molecular component ($a_w$ and $m_w$) and run the simulations for $2 \, \mathrm{Gyr}$. In this work, we focus on the simulation assuming that star formation is only linked to the molecular fraction ($m_w=1$, $a_w=0$). 

\begin{figure}[!ht]
    \centering
    \includegraphics[width=0.5\columnwidth]{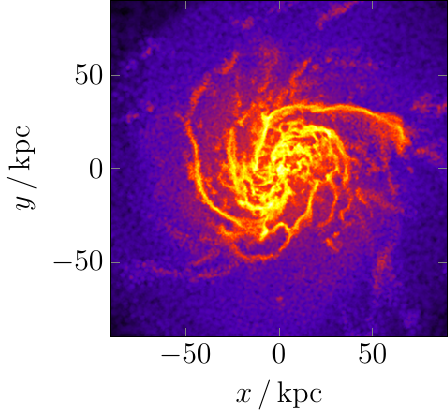}\includegraphics[width=0.5\columnwidth]{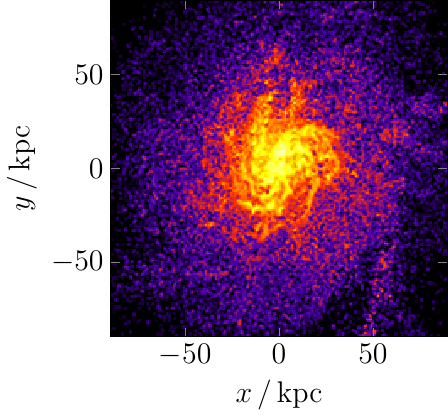}
    \caption{Projected gas (\emph{left panel}) and molecular (\emph{right panel}) densities for our simulation after $2 \, \mathrm{Gyr}$ of evolution, in a face-on view.}
    \label{fig:density_map}
\end{figure}

Fig.~\ref{fig:density_map} shows the projected total and molecular gas densities, in a face-on view, for our simulation at the final snapshot. The gas is well-settled into a rotationally-supported structure reminiscent of a spiral galaxy, with the usual complexity of the hydrodynamical gas evolution. As expected, the molecular gas does not follow exactly the distribution of the total gas, as its formation depends not only on the gas density but also on the gas metallicity. 
It is important to note that it is the molecular hydrogen that, in our model, is actively participating in the process of forming new stars, implying that the stellar spatial distribution will be determined by the location of molecular clouds.

\begin{figure}[!htbp]
    \centering
    \includegraphics[width=0.9\columnwidth]{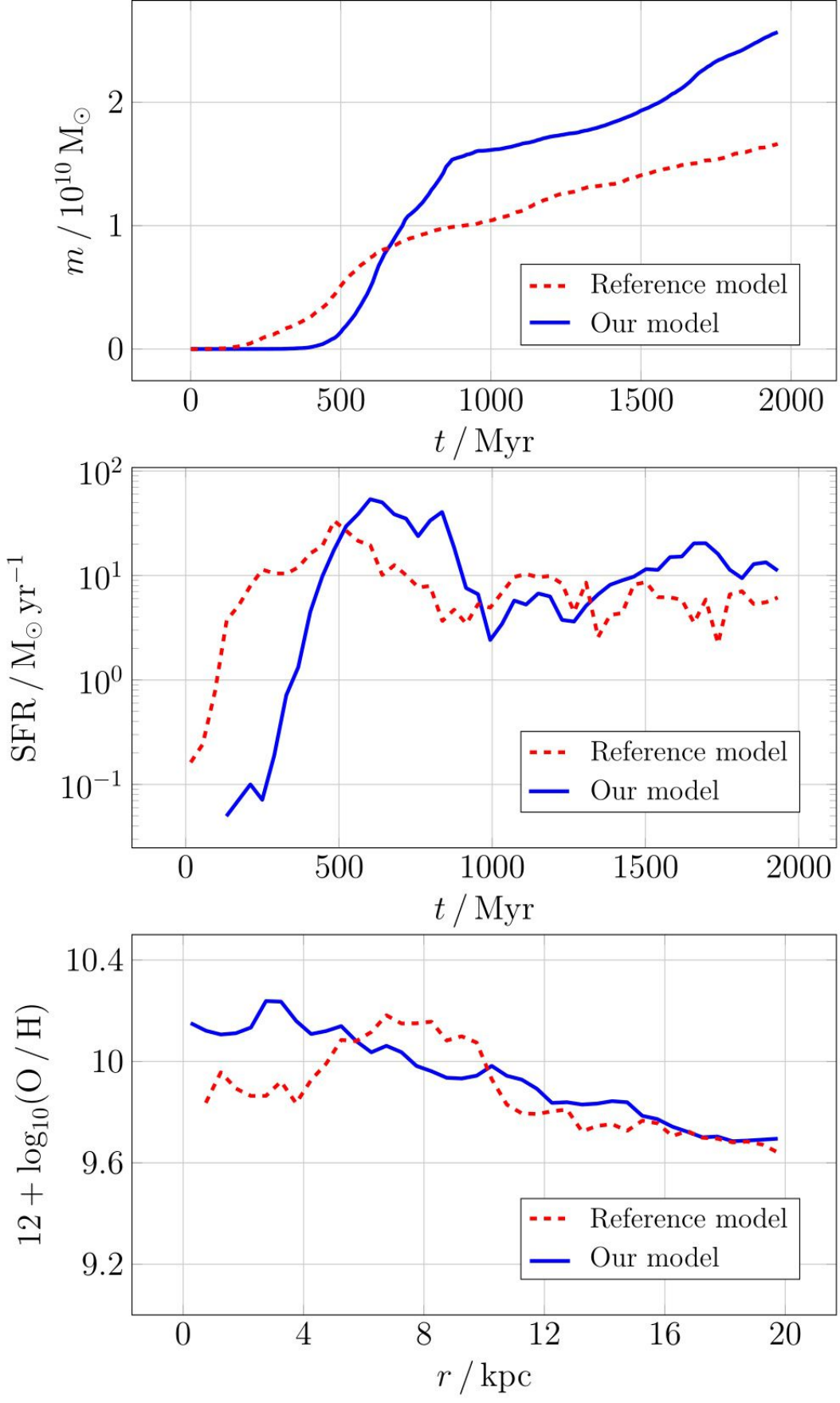}
    \caption{Integrated stellar mass (\emph{top panel}) and star formation rate (\emph{middle panel}) of our simulation (solid blue), as a function of time, compared to results of a simulation where star formation is linked to the total gas density (dashed red). The \emph{bottom panel} shows the corresponding gas oxygen abundance profile at the end of the simulations.}
    \label{fig:mass_vs_time}
\end{figure}

Fig.~\ref{fig:mass_vs_time} compares the results of our simulation (solid line in blue) and a reference model (dashed line in red) in which the star formation rate is solely determined by the total gas density as in standard implementations.
The upper two panels of this figure show the evolution of the integrated stellar mass and SFR. The most important effect of linking the SFR to the molecular fraction is a delay in the onset of star formation  of around $200 \, \mathrm{Myr}$ with respect to the reference model.
This delay is a consequence of the evolution of the gas in the  molecular phase, which needs some time to be created and is regulated by the level of enrichment of the gas, determined, in turn, by star formation.  After this first period, when enough stars have formed and the interstellar gas is enriched, the SFR reaches higher values compared to the reference model, producing a significant rise in the total stellar mass formed. At the end of the simulation, our model produced a higher total stellar mass even though the stars started to form later compared to the reference model. 

The bottom panel of Fig.~\ref{fig:mass_vs_time} displays the oxygen abundance profile at the end of the simulations.  Although the two profiles are similar in the outer regions, our model has a more pronounced peak at the center of the disc, similar to observed profiles (similar results are obtained for other elements). 
The differences follow the variations in the distribution of molecular and total gas that determine the shape and levels of the metallicity profiles.

\section{Conclusions}
\label{sec:conclusions}

We presented a new model to describe the atomic and molecular fractions within a gas cloud, which is used to implement a star formation rate coupled to these fractions in arbitrary proportions. Our  model has been specifically designed to be used in conjunction with simulations of galaxy formation in a cosmological context. 

In order to test the effectiveness of our model, we applied it to simulations of isolated halos with a mass comparable to that of the Milky Way, and  compared our findings to a reference model where the star formation rate is determined solely by the total gas density. Our results indicate that the dependence of the star formation rate on the amount of molecular gas delays  the onset of star formation, and could affect the metallicity profiles in disc galaxies.

\begin{acknowledgement}
The authors acknowledge support provided by UBACyT 20020170100129BA
\end{acknowledgement}


\bibliographystyle{baaa}
\small
\bibliography{bibliografia}
 
\end{document}